\documentclass[twocolumn,showpacs,preprintnumbers,amsmath,amssymb]{revtex4}
\usepackage{graphicx}
\usepackage{dcolumn}
\usepackage{bm}

\def\be{\begin{equation}}
\def\ee{\end{equation}}
\def\ba{\begin{array}}
\def\ea{\end{array}}
\def\bea{\begin{eqnarray}}
\def\eea{\end{eqnarray}}

\begin{document}
\preprint{APS/123-QED}

\title{\large\bf Multifragmentation at the balance energy of mass asymmetric colliding nuclei.}
\author{Supriya Goyal}
\email{supriyagoyal.pu@gmail.com}
\affiliation{\it House No. 86, Phase-III, Urban Estate, Dugri
Road, Ludhiana, Punjab - 141 013, India\\ }
\date{\today}

\begin{abstract}
Using the quantum molecular dynamics model, we study the role of
mass asymmetry of colliding nuclei on the fragmentation at the
balance energy and on its mass dependence. The study is done by
keeping the total mass of the system fixed as 40, 80, 160, and 240
and by varying the mass asymmetry of the ($\eta$ =
$\frac{A_{T}-A_{P}}{A_{T}+A_{P}}$; where $A_{T}$ and $A_{P}$ are
the masses of the target and projectile, respectively) reaction
from 0.1 to 0.7. Our results clearly indicate a sizeable effect of
the mass asymmetry on the multiplicity of various fragments. The
mass asymmetry dependence of various fragments is found to
increase with increase in total system mass (except for heavy mass
fragments). Similar to symmetric reactions, a power law system
mass dependence of various fragment multiplicities is also found
to exit for large asymmetries.
\end{abstract}

\maketitle

\section{Introduction}

The main goals in the study of heavy-ion collisions at the
intermediate energies are the determination of the bulk properties
of nuclear matter, or the nuclear equation of state, and the
understanding of the collision processes, which vary over the
large range of energies available today. These goals are related
to each other and an improved insight into one can lead to a
better understanding of the other. The study of multifragmentation
in the intermediate energy range gives us a possibility to
understand the properties of nuclear matter at extreme conditions
of temperature and density. The detailed experimental and
theoretical studies clearly point towards the dependence of the
reaction dynamics on entrance channel parameters such as incident
energy, impact parameter as well as mass asymmetry of the
colliding nuclei \cite{1,1a,2,3,4,5}.
\par
It is well known that the reaction dynamics for symmetric and
asymmetric reactions are entirely different. The former leads to
higher compression whereas the latter has a large share as thermal
energy \cite{6}. In a recent study by Puri and collaboration, a
detailed analysis is presented on the effect of mass asymmetry of
colliding nuclei on the collective flow and its disappearance,
nuclear stopping, elliptical flow, multifragmentation (at fixed
energies), and nuclear dynamics (at the balance energy $E_{bal}$;
i.e. energy at which collective flow disappears) by keeping the
total mass of the system fixed and at different impact parameters
\cite{7,7a}. A sizeable role of mass asymmetry has been found in
all the cases. Unfortunately, the role of mass asymmetry of the
colliding nuclei on the fragment structure at the balance energy
is not presented in the literature. A similar attempt was made by
Dhawan and Puri \cite{8}, but it was limited for symmetric
colliding nuclei only. Therefore, in the present work, we aim to
see the effect of mass asymmetry of the colliding nuclei on the
fragment structure and its mass dependence by simulating the
reactions at their corresponding balance energies. The mass
asymmetry of the reaction is varied by keeping the total mass of
the system fixed. The quantum molecular dynamics (QMD) model
\cite{1,1a,2,4,5,6,7,7a,8,9,10} is used for the present study and
is explained in section II. Section III is devoted to the results
and discussion followed by summary in section IV.

\section{The Model}

In quantum molecular dynamics model
\cite{1,1a,2,4,5,6,7,7a,8,9,10}, nucleons (represented by Gaussian
wave packets) interact via mutual two- and three-body
interactions. Here each nucleon is represented by a coherent state
of the form:
\begin{equation}
\phi_{i}(\vec{r},\vec{p},t)=\frac{1}{\left(2\pi
L\right)^{3/4}}e^{\left[-\left\{\vec{r}-\vec{r}_{i}(t)\right\}^2/4L\right]}
e^{\left[i\vec{p}_{i}(t)\cdot\vec{r}/\hbar\right]}.
\end{equation}
The Wigner distribution of a system with ${\it A_{T}+A_{P}}$
nucleons is given by
\begin{equation}
f(\vec{r},\vec{p},t)=\sum_{i = 1}^{A_{T}+A_{P}}\frac{1}{\left(\pi
\hbar\right)^{3}}
e^{\left[-\left\{\vec{r}-\vec{r}_{i}(t)\right\}^2/2L\right]}
e^{\left[-\left\{\vec{p}-\vec{p}_{i}(t)\right\}^2
2L/\hbar^{2}\right]^{'}},
\end{equation}
with L = 1.08 $fm^{2}$.
\par
The center of each Gaussian (in the coordinate and momentum space)
is chosen by the Monte Carlo procedure. The momentum of nucleons
(in each nucleus) is chosen between zero and local Fermi momentum
[$=\sqrt{2m_{i}V_{i}(\vec{r})}$; $V_{i}(\vec{r})$ is the potential
energy of nucleon $i$]. Naturally, one has to take care that the
nuclei, thus generated, have right binding energy and proper root
mean square radii.
\par
The centroid of each wave packet is propagated using the classical
equations of motion:
\begin{equation}
\frac {d\vec{r}_{i}}{dt} = \frac {dH}{d\vec{p}_{i}},
\end{equation}
\begin{equation}
\frac {d\vec{p}_{i}}{dt} = -\frac {dH}{d\vec{r}_{i}},
\end{equation}
where the Hamiltonian is given by
\begin{equation}
H=\sum_{i} \frac {\vec{p}_{i}^{2}}{2m_{i}} + V ^{tot}.
\end{equation}
 Our total interaction potential $V^{tot}$ reads as
\begin{equation}
V^{tot} = V^{Loc} + V^{Yuk} + V^{Coul} + V^{MDI},
\end{equation}
with
\begin{equation}
V^{Loc} = t_{1}\delta(\vec{r}_{i}-\vec{r}_{j})+
t_{2}\delta(\vec{r}_{i}-\vec{r}_{j})
\delta(\vec{r}_{i}-\vec{r}_{k}),
\end{equation}
\begin{equation}
V^{Yuk}=t_{3}e^{-|\vec{r}_{i}-\vec{r}_{j}|/m}/\left(|\vec{r}_{i}-\vec{r}_{j}|/m\right),
\end{equation}
with ${\it m}$ = 1.5 fm and $\it{t_{3}}$ = -6.66 MeV.
\par
The static (local) Skyrme interaction \cite{11} can further be
parametrized as:
\begin{equation}
U^{Loc}=\alpha\left(\frac{\rho}{\rho}_o\right)+
\beta\left(\frac{\rho}{\rho}_o\right)^{\gamma}.
\end{equation}
Here $\alpha, \beta$ and $\gamma$ are the parameters that define
equation of state. The momentum dependent interaction is obtained
by parameterizing the momentum dependence of the real part of the
optical potential. The final form of the potential reads as
\begin{equation}
U^{MDI}\approx t_{4}ln^{2}[t_{5}({\it\vec{p}_{i}}-{\it
\vec{p}_{j}})^{2}+1]\delta({\it \vec{r}_{i}}-{\it \vec{r}_{j}}).
\end{equation}
Here ${\it t_{4}}$ = 1.57 MeV and ${\it t_{5}}$ = 5 $\times
10^{-4} MeV^{-2}$. A parameterized form of the local plus momentum
dependent interaction (MDI) potential (at zero temperature) is
given by
\begin{equation}
U=\alpha \left({\frac {\rho}{\rho_{0}}}\right) + \beta
\left({\frac {\rho}{\rho_{0}}}\right)+ \delta
ln^{2}[\epsilon(\rho/\rho_{0})^{2/3}+1]\rho/\rho_{0}.
\end{equation}
The parameters $\alpha$, $\beta$, and $\gamma$ in above Eq. (11)
must be readjusted in the presence of momentum dependent
interactions so as to reproduce the ground state properties of the
nuclear matter. The set of parameters corresponding to different
equations of state can be found in Ref. \cite{1}.

\section{Results and Discussion}

For the present work, we simulated the central reactions of
$^{17}_{8}O+^{23}_{11}Na$ ($\eta = 0.1$),
$^{14}_{7}N+^{26}_{12}Mg$ ($\eta = 0.3$),
$^{10}_{5}B+^{30}_{14}Si$ ($\eta = 0.5$), and
$^{6}_{3}Li+^{34}_{16}S$ ($\eta = 0.7$) for $A_{TOT}$ = 40,
$^{36}_{18}Ar+^{44}_{20}Ca$ ($\eta = 0.1$),
$^{28}_{14}Si+^{52}_{24}Cr$ ($\eta = 0.3$),
$^{20}_{10}Ne+^{60}_{28}Ni$ ($\eta = 0.5$), and
$^{10}_{5}B+^{70}_{32}Ge$ ($\eta = 0.7$) for $A_{TOT}$ = 80,
$^{70}_{32}Ge+^{90}_{40}Zr$ ($\eta = 0.1$),
$^{54}_{26}Fe+^{106}_{48}Cd$ ($\eta = 0.3$),
$^{40}_{20}Ca+^{120}_{52}Te$ ($\eta = 0.5$), and
$^{24}_{12}Mg+^{136}_{58}Ce$ ($\eta = 0.7$) for $A_{TOT}$ = 160,
and $^{108}_{48}Cd+^{132}_{56}Ba$ ($\eta = 0.1$),
$^{84}_{38}Sr+^{156}_{66}Dy$ ($\eta = 0.3$),
$^{60}_{28}Ni+^{180}_{74}W$ ($\eta = 0.5$), and
$^{36}_{18}Ar+^{204}_{82}Pb$ ($\eta = 0.7$) for $A_{TOT}$ = 240,
at their corresponding theoretical balance energies (taken from
Ref. \cite{7}). The balance energies at which these reactions are
simulated were calculated using a momentum dependent soft equation
of state with standard energy dependent cugnon cross-section. The
reactions are followed uniformly up to 500 fm/c. A simple spatial
clusterization algorithm dubbed as minimum spanning tree (MST)
method is used to clusterize the phase space \cite{1}.
\begin{figure}
  \vskip -1.0cm
\includegraphics[width = 8 cm]{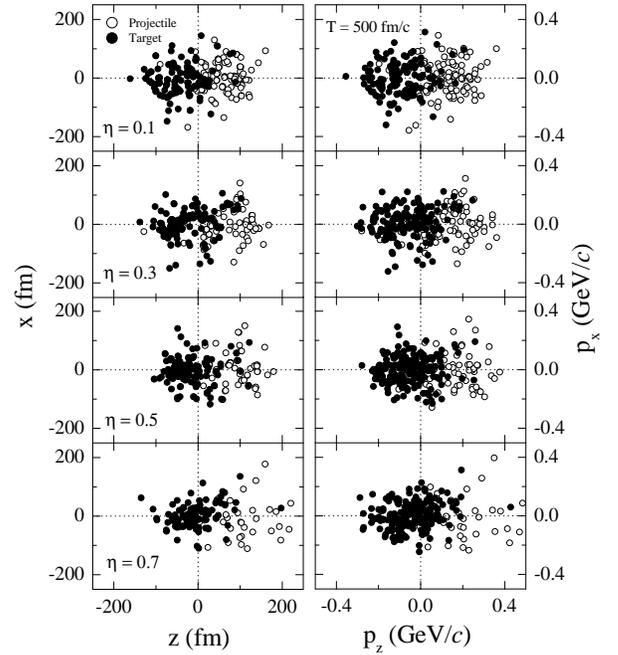}\\
\vskip - 0.5 cm
 \caption{Snapshots of a single event in the
phase space (x,z), left side, and $(p_{x},p_{z})$, right side, for
fixed system mass $A_{TOT}$ = 240 and $\eta$ = 0.1, 0.3, 0.5, and
0.7 at their corresponding balance energies.} \label{fig.1}
\end{figure}
\par
In Fig. 1, we display the snapshots of the final phase-space
(i.e.; $x$-$z$ (left column) and $p_{x}$-$p_{z}$ (right column))
of a single event at the balance energy for $\eta$ = 0.1-0.7 by
keeping the total mass of the system fixed as $A_{TOT}$ = 240. We
see a isotropic emission of nucleons for nearly symmetric
colliding nuclei whereas a binary character starts emerging out as
$\eta$ increases. One can say that phase-space is less homogenous
for large asymmetries. The behavior is similar in spatial and
momentum spaces. The above picture is quite similar for large
number of different events indicating a uniform distribution.
\begin{figure}
  \vskip -1.0cm
\includegraphics[width = 8 cm]{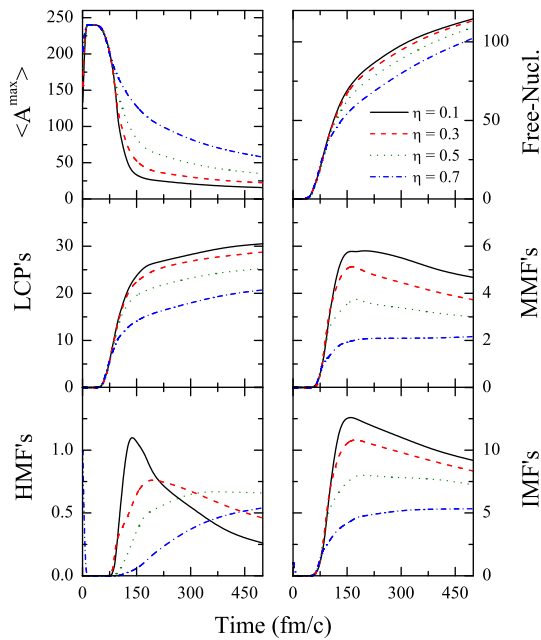}\\
\vskip - 0.5 cm
 \caption{(Color Online) Time evolution of the largest
fragment $A^{max}$, free nucleons, LCP's $(2\leq A \leq 4)$, MMF's
$(5\leq A \leq 9)$, HMF's $(15\%\leq A \leq 30\%)$ and IMF's
$(4\leq A \leq 30\%)$ for fixed system mass $A_{TOT}$ = 240 and
$\eta$ = 0.1, 0.3, 0.5, and 0.7 at their corresponding balance
energies. The solid, dashed, dotted, and dashed-dotted lines,
correspond to $\eta$ = 0.1, 0.3, 0.5, and 0.7, respectively.}
\label{fig.2}
\end{figure}
\par
In Fig. 2, we display the time evolution of the largest fragment
survived $A^{max}$, free nucleons, the light charged particles
(LCP's) $2\leq A \leq4$, the medium mass fragments (MMF's) $5\leq
A \leq 9$, the heavy mass fragments (HMF's) $15\%\leq A \leq 30\%$
as well as the intermediate mass fragments (IMF's) $4\leq A \leq
30\%$ (of the largest between target and projectile). The results
are displayed for different mass asymmetries by keeping the total
mass fixed as $A_{TOT}$ = 240. In order to avoid unwanted and
artificial heavy fragments for large asymmetries and lighter
colliding nuclei, the percentages are taken in HMF and IMF
definitions. As expected, $A^{max}$ has a peak around 20-80 fm/c
for all $\eta$. The excited compound nucleus formed in the early
stage is independent of $\eta$ and decays by the emission of
nucleons and fragments. Therefore, the free nucleons and LCP's
display a constant rise in their multiplicity. The MMF's, HMF's,
and IMF's are unstable and decay at a later time. The size of
largest fragment at final stage increases with increase in $\eta$
whereas a reverse trend is seen for free nucleons, LCP's, MMF's,
and IMF's. This is because of a decrease in the participant zone
and hence net nucleon-nucleon collisions, with increase in $\eta$.
The trend that HMF's follows with $\eta$ is different. It is clear
from the figure that the multiplicity of HMF's increases with time
for larger asymmetries. Also, since balance energy for large
asymmetric colliding nuclei is larger than that in small
asymmetric nuclei, it takes small time for large asymmetric
colliding nuclei to saturate.
\begin{figure}
  \vskip -1.0cm
\includegraphics[width = 8 cm]{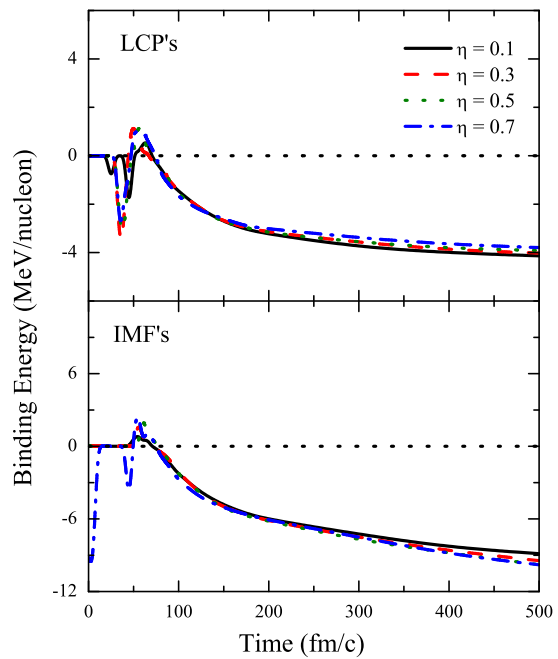}\\
\vskip - 0.5 cm
 \caption{(Color Online) The average binding energy per
nucleon of LCP's and IMF's as a function of time for fixed system
mass $A_{TOT}$ = 240 and $\eta$ = 0.1, 0.3, 0.5, and 0.7 at their
corresponding balance energies. Lines have same meaning as in Fig.
2.} \label{fig.3}
\end{figure}
\begin{figure}
  \vskip -1.0cm
\includegraphics[width = 8 cm]{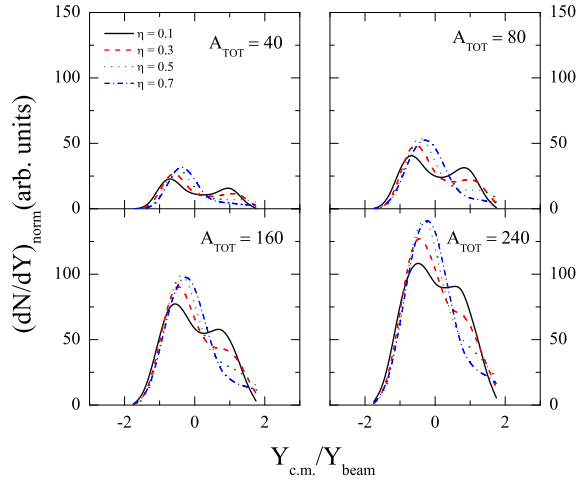}\\
\vskip - 2.5 cm
 \caption{(Color Online) Normalized rapidity distribution
$\frac{1}{A_{P}+A_{T}}$ $\frac{dN}{dY}$ as a function of scaled
rapidity $Y_{c.m.}/Y_{beam}$ for $\eta$ = 0.1-0.7 by keeping the
system mass fixed as $A_{TOT}$ = 40-240. Lines have same meaning
as in Fig. 2.} \label{fig.4}
\end{figure}
\begin{figure}
  \vskip -1.0cm
\includegraphics[width = 8 cm]{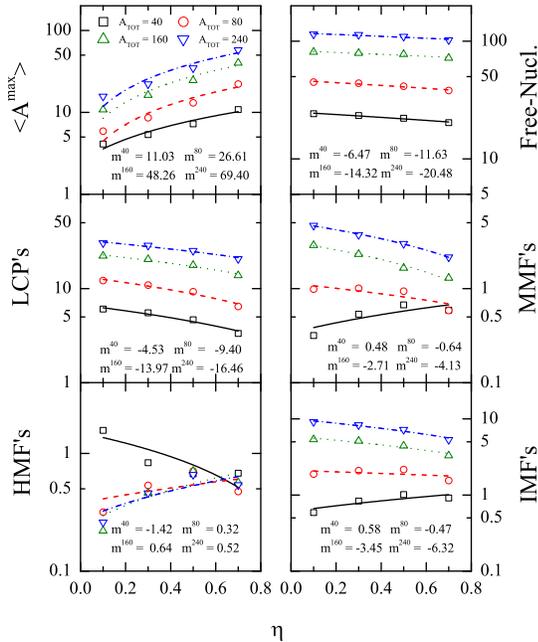}\\
\vskip - 0.5 cm
 \caption{(Color Online) The multiplicities of $A^{max}$, free
nucleons, LCP's, MMF's, HMF's and IMF's as a function of mass
asymmetry of colliding nuclei. The results for different system
masses $A_{TOT}$ = 40, 80, 160, and 240 are represented,
respectively, by the open squares, circles, triangles and inverted
triangles. Lines are the linear fits ($\propto m\eta$); {\it m}
values without errors are displayed.} \label{fig.5}
\end{figure}
\begin{figure}
  \vskip -1.0cm
\includegraphics[width = 8 cm]{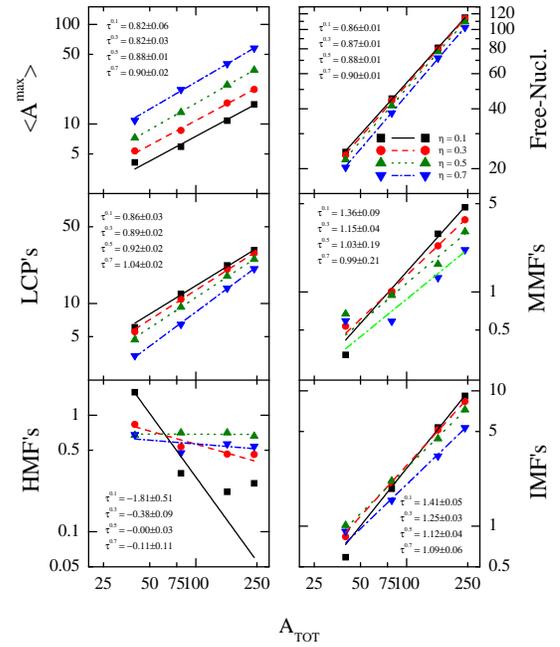}\\
\vskip - 0.5 cm
 \caption{(Color Online) Same as Fig. 5, but as a function of
total mass of the system. The results for different asymmetries
$\eta$ = 0.1, 0.3, 0.5, and 0.7 are represented, respectively, by
the solid squares, circles, triangles and inverted triangles. The
lines are power law ($\propto A_{TOT}^{\tau}$) fits to the
calculated results. The values of the power factor $\tau$ are
displayed in the figure for various quantities.} \label{fig.6}
\end{figure}
\par
In Fig. 3, we display the time evolution of average binding energy
per nucleon of LCP's and IMF's for $\eta$ = 0.1-0.7 by keeping the
total mass fixed as 240. It is clear from the figure that
independent of $\eta$, all fragments are reasonably bound. The
average binding energy/nucleon is around -4 MeV/nucleon for LCP's,
whereas it increases to around -8 MeV/nucleon for IMF's.
\par
In Fig. 4, we display the normalized rapidity distribution
((dN/dY)$_{norm}$) as a function of scaled rapidity
($Y_{c.m.}/Y_{beam}$) for $\eta$ = 0.1-0.7 by keeping the system
mass fixed as $A_{TOT}$ = 40-240. The rapidity is defined as:
\begin{equation}
Y(j)= \frac{1}{2}\ln\frac{{\bf{E}}(j)+{\bf{p}}_{z}(j)}
{{\bf{E}}(j)-{\bf{p}}_{z}(j)}.
\end{equation}
Here ${{\bf E}(j)} $ and $ {\bf p}_{z}(j)$ are, respectively, the
total energy (nucleon) and longitudinal momentum per nucleon for
the $jth$ nucleon. The parameter $Y_{c.m.}/Y_{beam}$ = 0
corresponds to the mid-rapidity (participant) zone and, hence, is
responsible for the hot and compressed zone. On the other hand,
$Y_{c.m.}/Y_{beam}\neq 0$ corresponds to the spectator zone,
($Y_{c.m.}/Y_{beam}<$ -1 corresponds to target-like (TL) and
$Y_{c.m.}/Y_{beam}>$ 1 corresponds to projectile-like (PL)
distributions). We see that the rapidities of nucleons emitted for
$\eta$ = 0.1-0.7 are not similar. Due to large balance energy for
larger asymmetries, single broader Gaussian is observed that is
peaked around the target rapidity, as the major contribution is
due to the target in all cases. As mass asymmetry decreases, the
balance energy decreases, therefore, one find peaks at target and
projectile rapidities indicating a non-equilibrium situation.
However, if the reactions would have been simulated at a fixed
incident energy, the peak shifts toward the mid-rapidity with the
decrease of the mass asymmetry and a greater thermalization would
have been observed in the case of a nearly symmetric collision
compared to an asymmetric collision. The same trend is seen for
all fixed system masses.
\par
In Fig. 5, we display the mass asymmetry dependence of different
fragments shown in Fig. 2 for $A_{TOT}$ = 40-240. Lines are the
linear fits ($\propto m\eta$). The values of $m$ are displayed in
figure. The mass of the largest fragment increases with increase
in $\eta$ for each $A_{TOT}$, whereas an opposite trend is seen
for free nucleons, LCP's, MMF's (except $A_{TOT}$ = 40), and IMF's
(except $A_{TOT}$ = 40). The multiplicity of HMF's show entirely
different behavior. It is clear from the figure that $\eta$
dependence increases with increase in system mass. This is because
of decrease in balance energy with increase in $A_{TOT}$. At low
incident energies, the Pauli-principal hinders the nucleon-nucleon
collisions and the increase of mass asymmetry further adds the
same effect. While at large incident energies for smaller
$A_{TOT}$, the role of $\eta$ decreases compared to large system
masses.
\par
Similar to Fig. 5, we display the mass dependence of various
fragments in Fig. 6. The mass asymmetry of the reaction is varied
from 0.1 to 0.7. Lines are power law fits ($\propto
A_{TOT}^{\tau}$); where values of power factor $\tau$ are
displayed in the figure. Similar to mass symmetric reactions
\cite{8}, a power law system mass dependence for various fragment
multiplicities exits for larger asymmetries. All the quantities
except HMF's show increasing trends for each $\eta$. It is clear
from the values of $\tau$ that, for $A^{max}$, free nucleons, and
LCP's; the mass dependence increases with increase in $\eta$,
whereas opposite trend is seen for MMF's, HMF's, and IMF's. The
trend of MMF's, HMF's and IMF's with change in $\eta$ from 0.1 to
0.7 in the lighter mass range gets reversed as one goes to higher
mass range. This is because for lighter system mass, the incident
energy is large compared to heavier system mass, therefore, large
mass asymmetric colliding nuclei will produce more heavy
fragments. The situation is entirely opposite for heavier system
mass.

\section{Summary}

We presented the study of role of mass asymmetry of colliding
nuclei on the fragmentation at the balance energy and on its mass
dependence using quantum molecular dynamics model. The analysis
was done by keeping the total mass of the system fixed as 40, 80,
160, and 240 and by varying the mass asymmetry of the colliding
nuclei from 0.1 to 0.7. We find a sizeable effect of the mass
asymmetry on the multiplicity of various fragments. Our finding at
the balance energy clearly point towards a power law system mass
dependence of different fragment multiplicities for each mass
asymmetric colliding nuclei..\\
\\
Author is thankful to Dr. Rajeev K. Puri for interesting and
constructive discussions. This work is supported by a research
grant from the Council of Scientific and Industrial Research
(CSIR), Govt. of India, vide grant no. 09/135(0563)/2009-EMR-1.


\end{document}